\documentstyle[12pt]{article}

\advance \textheight by \topskip
\begin{document}

\begin{flushright}
{\bf hep-th/9511137}
\end{flushright}
$\ $
\vskip 2truecm

\begin{center}

{ \large \bf  ANYONS AS SPIN PARTICLES:}\\[0.3cm]
{ \large \bf FROM CLASSICAL MECHANICS
TO FIELD THEORY}\footnote{Talk given at the Conference
``Theories of Fundamental Interactions", Maynooth, Ireland, May 1995;
to be published in ``Topics in Quantum Field Theory"
(World Scientific, Ed. T. Thcrakian, 1995)}\\
\vskip1.5cm
{ \bf Mikhail S. Plyushchay\footnote{On leave from the
{\it Institute for High Energy Physics,
Protvino, Moscow Region, Russia};
e-mail: mikhail@posta.unizar.es}}\\[0.5cm]
{\it Departamento de F\'{\i}sica Te\'orica,
Facultad de Ciencias}\\
{\it Universidad de Zaragoza, 50009 Zaragoza, Spain}\\[0.5cm]
\end{center}

\vskip2.0cm
\begin{center}
                            {\bf Abstract}
\end{center}
(2+1)-dimensional relativistic fractional spin particles are
considered within the framework of the group-theoretical
approach  to anyons starting from the level of classical
mechanics and concluding by the construction of the minimal set
of linear differential field equations.

\makeatletter
\@addtoreset{equation}{section}
  \def\theequation{\thesection.\arabic{equation}}
\makeatother

\newpage
\section{Introduction}
The case of (2+1)-dimensional space-time is special from
the point of view of spin and statistics: unlike the higher
dimensional cases, here spin of particles can take arbitrary
values on the real line and statistics interpolating
Fermi and Bose statistics is possible.  The first
peculiarity is coded in the topology of the (2+1)-dimensional
Lorentz group and its rotational subgroup manifolds,
$\pi_1(SO(2,1))=\pi_1(SO(2))=Z$ \cite{bin}, whereas the
possibility of exotic statistics is realizable due to a
nontrivial nature of the fundamental group of the configuration
space of $N$ identical particles on the plane:
$\pi_1(C_N)=B_N$ \cite{lei,braid}.  Here $Z$ is an additive group of
integer numbers and $B_N$ is a braid group.

The dynamical mechanism, realizing these possibilities at the
quantum mechanical level, was proposed by Wilczek
\cite{wil,braid}. It consists in supplying the system of the
charged particles with (singular) magnetic fluxes being
concentrated at the positions of the particles.  This mechanism
is based on the Aharonov-Bohm effect, and, so, it is rooted in
the profound principle of non-locality in quantum theory.

Nonlocal nature of fractional (arbitrary) spin particles obeying
exotic statistics (anyons) also reveals itself at the level of
the quantum field theory, where anyons can be realized with the
help of the universal construction, which consists in minimal
coupling the Chern-Simons U(1) gauge field to the conserved
matter current \cite{braid,sem}. The initial formulation of the
theory with Chern-Simons gauge field has a local character, but
the operators advanced to represent anyons turn out to be
realizable as nonlocal operators given on the one-dimensional
path -- nonobservable half-infinite `string' going to the space
infinity \cite{banerj}.  The spin-statistics relation takes
place for these anyonic objects \cite{banerj,fro}, but it is not
clear how the theory generally is reduced to the local theory of
the usual integer and half-integer spin fields in the case when
spin of the nonlocal anyonic fields takes integer and
half-integer values.

One can try to describe anyons in an alternative way, as
we describe spin $s=n/2$ fields, $n=0,1,2,...,$
not using Chern-Simons U(1) gauge field constructions.
But a priori it is clear that such a group-theoretical
description of anyons cannot be realized
starting from the pseudoclassical mechanics \cite{pseud}. This
`no-go theorem' is rooted in the fact that the
quantization of Grassmann variables, which are used there for
taking into account spin degrees of freedom, leads to
finite-dimensional representations of the (2+1)-dimensional
Lorentz group, and, as a result, only integer and half-integer
spin fields can be described in this way \cite{cor1}.
Nevertheless, the non-Grassmannian approach
can be formulated on the
basis of pseudoclassical mechanics \cite{pl1}.
This approach uses
usual commuting translation invariant variables instead of
anticommuting Grassmann variables and
allows us to describe massless \cite{pl2} and massive \cite{pl3}
particles of arbitrary (fixed) integer and half-integer spin
and to realize spin systems with global supersymmetry \cite{pl4},
and, moreover, it naturally leads to the  group-theoretical
approach for fractional spin particles \cite{pl5,pl6}.

The purpose of the present paper is to review some recent
results obtained within the framework of the group-theoretical
approach to anyons as spin particles.

The paper is organized as follows.  Section 2 is devoted to the
consideration of the classical mechanics of fractional spin
particles. Here we construct the `minimal' classical model of
relativistic particle with arbitrary spin, which is quantized in
section 3. In section 3 we discuss also the `universal' vector
system of linear differential equations and construct the
minimal spinor system of linear differential equations. The
latter construction is realized with the help of the deformed
(extended) Heisenberg-Weyl algebra.  The last section contains
concluding remarks.

\section{Classical mechanics of fractional spin particles}

In 2+1 dimensions, spin $S=p^{\mu}{\cal J}_{\mu}/\sqrt{-p^{2}}$
has a pseudoscalar nature and, so, relativistic particle with
fixed mass $-p^2 =m^2$ and nonzero spin $S=s\neq 0$ has the
same number of degrees of freedom as a massive scalar ($s=0$)
particle. Therefore, one can try to describe spin particle
by the minimal set of phase space variables being the
coordinates $x_\mu$ and canonically conjugate energy-momentum vector
$p^\mu$, $\{x_\mu,p_\nu\}=\eta_{\mu\nu}$.
Then the total angular momentum vector of the system
can be chosen in the form
\[
{\cal J}_\mu
=-\epsilon_{\mu\nu\lambda}x^{\nu}p^{\lambda} +j_{\mu},
\]
generalizing the form of the total angular momentum vector of
the scalar particle \cite{corpl}. Here ${\cal J}_\mu$
is a vector dual to the total angular momentum tensor,
${\cal J}_\mu=-\frac{1}{2}\epsilon_{\mu\nu\lambda}{\cal
J}^{\nu\lambda}$, the second term
$j_{\mu}=j_{\mu}(p)$ serves for taking into account
a nontrivial spin $s$, $j_{\mu}=-sp^{\mu}/{\sqrt{-p^2}}+
j^{\bot}_{\mu}(p),$ $j^{\bot}p=0$, and we assume that $p^{2}$
is fixed with the help of the mass shell constraint,
\begin{equation}
\phi=p^2+m^2\approx 0.
\label{mass}
\end{equation}
Generally, the coordinates $x_\mu$ can have the Poisson brackets
of the form
\[
\{x_\mu,x_\nu\}=\epsilon_{\mu\nu\lambda}R^\lambda.
\]
The Jacobi identities for the brackets of $x_\mu$ and $p_\nu$ and
classical Poincar\'e algebra,
\[
\{p_\mu,p_\nu\}=0,\quad
\{{\cal J}_\mu,{\cal J}_\nu\}
=-\epsilon_{\mu\nu\lambda}{\cal J}^{\lambda},\quad
\{{\cal J}_\mu,p_\nu\}=-\epsilon_{\mu\nu\lambda}p^{\lambda},
\]
lead to the following most general form of the quantities
$R^{\mu}$ and $j^{\bot}_\mu$, which turn out to be correlated:
\[
R_\mu=R^c_\mu+
\epsilon_{\mu\nu\lambda}\partial^{\nu}A^{\lambda},\quad
R^{c}_\mu=-sp_{\mu}/(-p^2)^{3/2};\qquad
j^{\bot}_\mu= -\epsilon_{\mu\nu\lambda}
A^{\nu}p^{\lambda},
\]
where
$A_\mu=A_\mu(p)$ is a set of arbitrary functions having the
dimensionality of inverse mass.  Only in one special case, when
$A_\mu=p_\mu\cdot a(p^2)$ and, therefore, $R_\mu=R^c_\mu$ and
$j^\bot_\mu=0$, the coordinates of the particle $x_\mu=x^c_\mu$
form the Lorentz vector:  $\{{\cal
J}_\mu,x^c_\nu\}=-\epsilon_{\mu\nu\lambda}x^{c\lambda}$,
whereas in all other cases $x_\mu$ is not a Lorentz covariant
object.
The Poisson brackets
$\{x^c_\mu,x^c_\nu\}=\epsilon_{\mu\nu\lambda}R^{c\lambda}$
mean that at the quantum level there is no representation
where the corresponding operators of the coordinates of the
particle would be diagonal, and, hence, the {\it special
covariant case is characterized by nonlocalizable
coordinates}.

There is another special case characterized by localizable
coordinates $x_\mu=x_\mu^l$, $\{x^l_\mu,x^l_\nu\}=0$.
In this case the ``gauge field" $A_\mu$ is defined by the
equation $\partial_\mu A_\nu-\partial_\nu A_\mu=
\epsilon_{\mu\nu\lambda}R^{c\lambda}$ meaning that $A_\mu$
has a curvature of SO(2,1) monopole. It is given by the
expression
\[
A^l_\mu=\epsilon_{\mu 0i}R^{ci}\cdot f(p),
\qquad
f(p)=-\frac{p^0\cdot p^2}{\vert p^0\vert\cdot(\vert
p^0\vert+\sqrt{-p^2})},
\]
which can be considered as a most
general solution of the `curvature equation' \cite{corpl}.
One can convinced
that in correspondence with general properties, $x^l_\mu$ has
complicated transformation properties with respect to the pure
Lorentz transformations (boosts), and, hence, is not a Lorentz
vector. But due to the property of localizability, these
coordinates help to realize covariant operators $X^c_\mu$
corresponding to classical coordinates $x^c_\mu$:
\[
X^c_\mu=X^l_\mu+A^l_\mu(P).
\]
We shall return to the discussion of the quantization of
the described formulation of relativistic fractional
spin particle, which can naturally be called a {\it minimal
canonical formulation}, in the last section. So, we conclude
that within the framework of the minimal formulation the
properties of covariance and localizability for the coordinates
of arbitrary spin particle cannot be simultaneously
incorporated into the theory.

But it can be done via extending the phase space of the system
by `internal' phase space variables $z_n$, $n=1,\ldots,2N$,
$\{z_n,x_\mu\}=\{z_n,p_\mu\}=0$, and  supposing that spin
addition depends on these variables:
$j_\mu=j_\mu(z_n,p_\nu)$. In this case spin can be fixed by
imposing the spin constraint
\begin{equation}
\chi=pj-sm\approx 0.
\label{spin}
\end{equation}
Constraints (\ref{mass}) and (\ref{spin}) form
the set of first class constraints of the {\it extended
formulation}, whereas $N-1$ degrees of freedom, different
from the spin one, should be `frozen' by introducing the
corresponding number of first or/and second class
constraints.

Let us suppose now that $j_\mu$  does not depend on $p_\mu$, i.e.
$j_\mu=j_\mu(z_n)$, and that coordinates $x_\mu$ are
localizable, $\{x_\mu, x_\nu\}=0$. Then it follows that $j_\mu$
itself has to satisfy (2+1)-dimensional Lorentz algebra,
\begin{equation}
\{j_\mu,j_\nu\}=-\epsilon_{\mu\nu\lambda}j^{\lambda},
\label{jjj}
\end{equation}
and that $x_\mu$ is a Lorentz vector.
Therefore, within the framework of the extended formulation we
indeed can simultaneously incorporate into the theory the
properties of localizability and covariance of the coordinates.

The minimal case of extended formulation is
characterized by two internal phase space variables (one degree
of freedom frozen by the spin constraint),
and it can be realized in the following way.
First we note that $j^2$ lies in the centre of Lorentz
algebra (\ref{jjj}), and, therefore, it can be fixed by putting
$j^2=C=const$. As a result, we can consider $j_\mu$ subject to
this condition as the internal variables themselves. Then the
topology of the internal phase subspace will be defined by this
constant parameter $C$. In the case $C=-\alpha^2<0$,
$\alpha^2\leq s^2$, we have the two sheet hyperboloid,
$j_0=\pm\sqrt{\alpha^2+j_1^2+j_2^2}$, as the internal phase
subspace, whereas the case $C=\beta^2\geq 0$ gives a one sheet
hyperboloid.
The Lagrangian leading to brackets (\ref{jjj}) and
constraints (\ref{mass}) and (\ref{spin}), has the following form
\cite{corpl}:
\[
L_0=\frac{1}{2e}(\dot{x}_\mu-vj_\mu)^2
-\frac{e}{2}m^2+smv
-\frac{j\xi}{j^2+(j\xi)^2}\epsilon_{\mu\nu\lambda}
\xi^{\mu}j^{\nu}\dot{j}{}^{\lambda},
\]
where $\xi^\mu$ is a constant timelike vector, $\xi^2=-1$, and
$e$ and $v$ are Lagrange multipliers.

On the phase space of the described minimal extended system,
one can introduce the vector $\tilde{x}_\mu=
x_\mu+(p^2)^{-1}\epsilon_{\mu\nu\lambda}p^\nu j^\lambda$.
This vector has, unlike $x_\mu$,
zero brackets, $\{\tilde{x}_\mu,\tilde{\chi}\}=0$,
with the spin constraint $\tilde{\chi}$
presented with the
help of the mass shell constraint in the dimensionless form
$\tilde{\chi}= jp/\sqrt{-p^2} -s$.  So,
$\tilde{x}_\mu$ is a gauge invariant extension of the initial
coordinates $x_\mu$. It has a simple relativistic evolution
law, $d\tilde{x}{}^i/d\tilde{x}{}^0=p^i/p^0$.
On the other hand, the initial
coordinates  $x_\mu$ reveal more complicated motion
as a consequence of a classical analog of the quantum
relativistic Zitterbewegung \cite{pl1},
which generally takes place in the system \cite{corpl}.
In this respect the
coordinates $\tilde{x}{}_\mu$ are analogous to the
Foldy-Wouthuysen coordinates for the Dirac particle.
But there is one special case here, which
is characterized by the value of the parameter $C$ being
correlated  with the value of the spin parameter: iff
$-j^2=\alpha^2=s^2$, there is Lagrange constraint in the
system: $\dot{x}^2-(\dot{x}j)^2/j^2=0$.
This constraint means that the
velocity of the particle is parallel to the vector $j_\mu$ and,
as a consequence of the spin constraint, to the energy-momentum
vector $p_\mu$. So, in this special case ${x}_\mu$ has
the same evolution law as $\tilde{x}_\mu$, and, moreover, here
the gauge-invariant extension $\tilde{x}_\mu$ coincides with
$x_\mu$.  We shall see below that this case turns out to be
special also from the point of view of linear differential equations.
In conclusion of classical considerations, we note that
the Poisson brackets of the quantities $\tilde{x}_\mu$ have the
same form (on the surface of constraint (\ref{spin}) ) as the
Poisson brackets of the covariant coordinates $x^c_\mu$ in the
minimal formulation, whereas the general case of the minimal
formulation can be obtained from the extended formulation via
reduction of the extended system to the surface of the
spin constraint \cite{corpl}.

\section{Quantum theory}

In correspondence with classical relations (\ref{jjj}),
the operators $J_\mu$ must satisfy the algebra
of (2+1)-dimensional Lorentz group SO(2,1) (or SL(2,R) group
locally isomorphic to it):
\begin{equation}
[J_\mu,J_\nu]=-i\epsilon_{\mu\nu\lambda}J^{\lambda},
\label{}
\end{equation}
whereas the first class constraints turn into equations
\begin{equation}
(P^2+m^2)\Psi=0,\qquad
(PJ-sm)\Psi=0.
\label{klemaj}
\end{equation}

In correspondence with classical picture, at the quantum level
we also have two different cases. In the first case,
when $C=-\alpha^2<0$, $\alpha>0$, the quantization of
the variables $j_\mu$ leads to unitary irreducible representations
(UIRs) of the discrete type series
$D^{\pm}_{\alpha}$ of the group $\overline{\rm SL(2,R)}$
being the universal covering group of SL(2,R). In these
representations the Casimir operator takes value
$J^2=-\alpha(\alpha-1)$ substituting the corresponding
classical value of the constant $C$, and the
operator $J_0$ takes the eigenvalues $j_0=\pm(\alpha+n)$,
$n=0,1,\ldots$, i.e. here we have infinite-dimensional
half-bounded representations. In the case when $C=\beta^2\geq
0$, the quantization leads to the UIRs of the principal
continuous series $C^\vartheta_\sigma$, $\vartheta\in[0,1)$,
$J^2=\sigma=\beta^2+1/4$, and $j_0=\vartheta+n$, $n=0,\pm 1,\pm
2,\ldots$, (or of supplementary continuous series with
$0<\sigma<1/4$ \cite{corpl}), i.e.  in this case we have
unbounded infinite-dimensional representations in
correspondence with classical picture.

Passing over to the rest frame system, ${\bf p}=0$,
one can check that equations (\ref{klemaj}) have nontrivial
solutions under the coordinated choice of the representation
of $\overline{\rm SL(2,R)}$ and of the value of the spin
parameter $s$.  In particular, in the case when $\Psi$ carries
the representation of the discrete type series
$D^{\pm}_\alpha$, and $s=\epsilon\alpha$, $\epsilon=+1$ or
$-1$, (that corresponds to the special classical case mentioned
in the end of the preceding section), eqs. (\ref{klemaj}) have
nontrivial solution describing  the state with mass $m$ and
spin $s=\epsilon\alpha$ \cite{pl6}.

Considering equations (\ref{klemaj}) as field equations,
one could try to construct the corresponding field action
and then realize the secondary quantization of the theory
in order to reveal a spin-statistics relation for fractional
spin fields. But, unlike the case of the Dirac equation
and equation for the topologically massive vector U(1) gauge field
\cite{jactem}, equations (\ref{klemaj}) are completely independent,
and, so, they are not very suitable for realizing such a program.
Therefore,
we arrive at the problem of constructing the set of linear
differential equations (by analogy with above mentioned equations),
from which equations (\ref{klemaj})
would appear as a consequence of integrability condition
(see also refs. \cite{jn1}-\cite{sor}).

Before going over to the consideration of this problem, let us
note that in the above mentioned
special case, when $s^2=\alpha^2$ and $J^\mu\in D^{\pm}_{\alpha}$,
the second equation from the set (\ref{klemaj}) is the (2+1)-dimensional
analog of the Majorana equation \cite{maj}.
This equation describes the quantum states of the model of
relativistic particle with torsion \cite{pl6,pol},
and like the Majorana equation itself \cite{maj}, has solutions
in the massive $(p^2<0)$, massless $(p^2=0)$ and tachyonic
sectors $(p^2>0)$. Moreover, in the massive sector it has the
following mass spectrum: $M_n=m\alpha/\vert s_n\vert$, $n=0,1,\ldots$,
where $s_n=\epsilon(\alpha+n)$ is the spin of states.
So, from the point of view of the Majorana equation, the role of
the Klein-Gordon equation consists in removing the tachyonic and
massless states and in singling out from the infinite tower of
states the state with highest mass and lowest spin modulus.

The following `universal' vector set of linear differential
equations for fractional spin field,
\begin{equation}
V_\mu\Psi=0,\quad V_\mu=\alpha
P_\mu-i\epsilon_{\mu\nu\lambda}P^{\nu}J^{\lambda} +\epsilon m
J_\mu, \quad \epsilon=\pm 1,
\label{vect}
\end{equation}
was
proposed in ref. \cite{cor1}.  This set of equations has
nontrivial solutions in the case of the choice of the two types
of irreducible representations of the group
$\overline{\rm SL(2,R)}$:  either unitary infinite-dimensional
representations of the discrete type series $D^{\pm}_{\alpha}$
($\alpha>0)$ , or $(2j+1)$-dimensional nonunitary
representations of the discrete type series $\tilde{D}_j$ in
the case when $\alpha=-j$, $0<2j\in Z$, and $J^2=-j(j+1)$.  In
the cases when $\alpha=-j=-1/2$ and $\alpha=-j=-1$, and when
corresponding 2-dimensional spinor and 3-dimensional vector
representations are chosen, the vector system of equations
(\ref{vect}) is reduced to one equation being the Dirac or
Jackiw-Templeton-Schonfeld equation \cite{jactem},
respectively. In all other cases any two of three equations
(\ref{vect}) can be chosen as a basic set of linear differential
equations (\ref{vect}) and all the set of three equations is
necessary only to have a manifestly covariant set of equations.
Equations (\ref{klemaj}) appear as a consequence of the basic
equations (\ref{vect}) and, as a result, vector system of equations
(\ref{vect}) describes massive fields carrying irreducible representations
of the (2+1)-dimensional Poincar\'e group $\overline{\rm ISO(2,1)}$
characterized by mass $M=m$ and spin $s=\epsilon\alpha$.
Moreover, the vector set of equations of the form (\ref{vect})
has the following interesting property:  it singles out itself
only half-bounded infinite-dimensional representations
$D^{\pm}_\alpha$ as suitable for the description of fractional
spin fields, and reject the use, for the purpose,
of unbounded representations
of the continuous series  \cite{cor1}.  So, the
vector set of equations (\ref{vect}) gives some link in the
description of fractional spin fields and usual integer and
half-integer spin fields, but it is not a minimal set of linear
differential equations for describing fractional spin fields.
The minimal set of equations, as we can conclude, must contain
the set of two equations, and if we want to have a manifestly
covariant formulation of the theory, we have to look for a
spinor set of linear differential equations.

Such a system of equations can be constructed with the help of
the deformed (extended) Heisenberg-Weyl algebra involving the
Klein operator \cite{pl7}.
This algebra is given by the following commutation
relation \cite{vas}:
\begin{equation}
[a^-,a^+]=1+\nu K,
\label{def}
\end{equation}
where $K$ is the Klein operator, defined, in turn, by the relations
$K^2=1$, $Ka^{\pm}+a^{\pm}K=0$, whereas $\nu\in {\bf R}$ is a deformation
parameter. The algebra (\ref{def}) has unitary representations
in the case when $\nu>-1$ \cite{pl7}. Let us define now the position
$Q$ and momentum $\Pi$ operators, $a^{\pm}=(Q\mp i\Pi)/\sqrt{2}$,
and note that in the coordinate representation $\Psi=\Psi(q)$,
$Q\Psi(q)=q\Psi(q)$,
the operator $\Pi$ can be realized as $\Pi=-i({d}/{dq}+K\cdot
{\nu}/{2q})$, whereas $K$ can be considered as a parity
operator, $K\Psi(q)=\Psi(-q)$.

Now, let us consider the set of operators
\[
L_1=Q,\quad L_2=\Pi,\qquad
J_0=\frac{1}{4}(a^+a^-+a^-a^+),\quad
J_{\pm}=J_1\mp iJ_2=\frac{1}{2}(a^{\pm})^2.
\]
They satisfy the superalgebra
\begin{equation}
[L_\alpha,L_\beta]_{{}_+}=4i(J\gamma)_{\alpha\beta},\quad
[J_\mu,J_\nu]=-i\epsilon_{\mu\nu\lambda}J^{\lambda}\quad
[J_\mu,L_\alpha]=\frac{1}{2}(\gamma_\mu)_\alpha{}^{\beta}L_\beta,
\label{comm}
\end{equation}
where $(\gamma^0)_\alpha{}^\beta=-(\sigma^2)_\alpha{}^\beta$,
$(\gamma^1)_\alpha{}^\beta=i(\sigma^1)_\alpha{}^\beta$,
$(\gamma^2)_\alpha{}^\beta=i(\sigma^3)_\alpha{}^\beta$.
Relations (\ref{comm}) mean that the operators $L_\alpha$ and $J_\mu$
are the generators of $osp(1\vert 2)$ superalgebra with
Casimir operator $C_{osp}=J_\mu J^\mu -\frac{i}{8}L^\alpha L_\alpha$
taking here the value $C_{osp}=(1-\nu^2)/16$.
Moreover, the first and third relations mean that $L_\alpha$
are `square root' operators from the SL(2,R)
generators $J_\mu$ and they are components of
(2+1)-dimensional spinor. Note, that on the subspaces of even,
$\Psi_+(q)=\Psi_+(-q)$, and odd, $\Psi_-(q)=-\Psi_-(-q)$,
functions the generators $J_\mu$ act in an irreducible way
realizing representations $D^{+}_{\alpha_\pm}$ with
$\alpha_+=\frac{1}{4}(1+\nu)>0$,
$\alpha_-=\alpha_+ +1/2$, respectively:
$J^2\Psi_\pm=-\alpha_\pm(\alpha_\pm-1)\Psi_\pm$.
Note also here that the representations of the series
$D^{-}_{\alpha_\pm}$ can be obtained in a simple way
by the substitution $J_0\rightarrow -J_0$,
$J_\pm\rightarrow -J_\mp$. Now, let us consider the spinor
set of equations:
\begin{equation}
S_\alpha \Psi=0,\quad
S_\alpha= L^\beta((P\gamma)_{\beta\alpha}+\epsilon
m\epsilon_{\beta\alpha}),
\label{spinor}
\end{equation}
where we suppose that $\Psi=\Psi(x,q)$, $P_\mu=-i\partial/
\partial x^\mu$ and $\epsilon_{\alpha\beta}=-\epsilon_{\beta\alpha}$
is a spinor metric tensor.
The following relation is valid on the subspace of even functions
$\Psi_+(x,q)=\Psi_+(x,-q)$ \cite{pl*}:
\[
(\gamma_\mu)^{\alpha\beta}L_\alpha S_\beta\Psi_+=
V_\mu\Psi_+,
\]
where $V_\mu$ is given by eq. (\ref{vect}). This means that on the
subspace of even functions $\Psi_+$ the spinor system of
equations (\ref{spinor}) has nontrivial solutions describing the
states of mass $M=m$ and spin
$s=\epsilon\alpha_+\neq 0$.  On the other hand, one can check that on
the subspace of odd functions this spinor set of equations has
no nontrivial solutions \cite{pl7}.

In conclusion let us point out on the hidden nonlocality
of the present constructions. Indeed, the use of half-bounded
infinite-dimensional representations for the construction
of linear differential equations (describing effectively
one-component field in a covariant way \cite{pl7}) could be
associated with the half-infinite nonobservable `string' of the
nonlocal anyonic field operators in the approach which uses the
statistical Chern-Simons U(1) gauge field \cite{banerj}.  In
the case of minimal spinor set of linear differential equations
the hidden nonlocality equivalently reveals itself in the
dependence of even functions $\Psi_+(x,q)= \Psi_+(x,-q)$ on
continuous additional variable $q\in{\bf R}$, that effectively
correspondence to giving the fractional spin field on some
half-inifinite `string'.

\section{Concluding remarks}
Starting from the minimal canonical formulation described in section 2,
one could try to construct the field action
leading to the Klein-Gordon equation being the quantum
analog of the only constraint (\ref{mass}) of the theory.
But the problem consists here in the absence of the
representation with the covariant operators $X^c_\mu$ to
be diagonal. One could use a representation with diagonal operators
$X^l_\mu$, but these coordinates
have  complicated transformation properties with respect
to the Lorentz transformations \cite{corpl}, and, therefore,
the theory will have manifestly noncovariant character. Moreover,
it is necessary to take into account a nontrivial behaviour of
the corresponding field $\Psi(x^l_\mu)$ with respect to the Lorentz
transformations \cite{jn1}.

So, because of these problems it seems more appropriate to
work within a framework of the extended formulation,
where the possibility to describe arbitrary spin fields is
achieved through the use of the infinite-dimensional representations
of the universal covering group of SL(2,R) group.
However, here we have still an open problem of constructing
the field action which would lead to the system of minimal spinor
set of linear differential equations.
In this case we have two equations for one (infinite-component
\cite{pl7}, or
depending on additional continuous argument $q$) basic
fractional spin field. Therefore, it is necessary to
introduce into the theory some auxiliary field(s) and
the problem is how to realize
the extension of the system in some minimal way.
Having such field action, we could realize the secondary
quantization of the theory to reveal a spin-statistics relation.
The hidden nonlocal nature of the present constructions, speculated
in the end of the preceding section,
can be considered as an
indication \cite{banerj,fro}
that a spin-statistics relation indeed can be revealed
for fractional spin fields in the group-theoretical approach.

$\ $

The work was supported by MEC-DGICYT.

\end{document}